\documentclass{article}
\usepackage{LaThuileFPSpro}
\usepackage{epsfig}
\newcommand{\be}{\begin{equation}}
\newcommand{\ee}{\end{equation}}
\newcommand{\GeV}{\rm{GeV}}
\begin{document}
\title{ 
DETERMINATION OF $V_{us}$: \\RECENT PROGRESSES FROM THEORY
}
\author{
Vittorio Lubicz \\ 
{\em Dip. di Fisica, Universit\`a di Roma Tre and INFN, Sez. di Roma III,}
\\ 
{\em Via della Vasca Navale 84, I-00146 Rome, Italy}
}
\maketitle

\baselineskip=11.6pt

\begin{abstract}
Recent experimental and theoretical results on kaon semileptonic decays have
significantly improved the determination of the CKM matrix element $V_{us}$.
After briefly summarizing the impact of the new experimental determinations,
I will concentrate in this talk on the theoretical progresses, coming in
particular from lattice QCD calculations. These results lead to the estimate
$|V_{us}|=0.2250 \pm 0.0021$, in good agreement with the expectation based
on the determination of $|V_{ud}|$ and the unitarity of the CKM matrix. 
\end{abstract}
\newpage
\section{Introduction}
\label{sec:intro}
The determination of the Cabibbo angle is of particular phenomenological and
theoretical interest since it provides at present the most stringent
unitarity test of the CKM matrix. This is expressed by the ``first row"
unitarity condition:
\be
|V_{ud}|^2 + |V_{us}|^2 + |V_{ub}|^2 =1 \,.
\label{eq:1strow}
\ee
Since $|V_{ub}|\sim 10^{-3}$, its contribution to Eq.~(\ref{eq:1strow})
can be safely neglected. 

The value of $|V_{ud}|$ is accurately determined from nuclear superallowed 
$0^+\to 0^+$ beta decays. An analysis of the results, based on nine
different nuclear transitions, leads to the very precise
estimate\cite{marciano}\footnote{After this talk, the estimate of $|V_{ud}|$
from nuclear superallowed decays has been updated by Marciano at the CKM
2005 Workshop on the Unitarity Triangle. The new estimate, whose
uncertainty is further reduced, reads $|V_{ud}| = 0.9739 \pm
0.0003$\cite{marciano2005}.} 
\be
\label{eq:vud}
|V_{ud}| = 0.9740 \pm 0.0005 \,.
\ee
This determination of $|V_{ud}|$ is more accurate, by approximately a factor
three, than the one obtained from the analysis of neutron beta decay. In
the neutron case, the error is dominated by the uncertainty on the
contribution of the weak axial current, which is determined experimentally,
$g_A/g_V=1.2720 \pm 0.0022$\cite{yellow}. It is also worth to mention that a
new measurement of the neutron lifetime has been recently
presented\cite{serebrov} whose new value, $\tau_n=(878.5 \pm 0.8)$ sec.,
differs by more than six standard deviations with respect to the previous
average quoted by the PDG, $\tau_n=(885.7 \pm 0.8)$ sec.\cite{PDG}. Combined
together, the neutron beta decay results lead to the determination $|V_{ud}|
= 0.9750 \pm 0.0017$, in agreement with Eq.~(\ref{eq:vud}) but with a much
larger error. In the following, I will take the estimate in
Eq.~(\ref{eq:vud}) as the final average of $|V_{ud}|$, and concentrate the
discussion on the remaining entry of Eq.~(\ref{eq:1strow}), the matrix
element $|V_{us}|$.

The most accurate determination of $|V_{us}|$ is obtained from semileptonic
kaon decays ($K_{\ell3}$). The analysis of the experimental data gives
access to the quantity $|V_{us}| \cdot f_+(0)$, where $f_+(0)$ is the vector
form factor at zero four-momentum transfer square. In the SU(3) limit,
vector current conservation implies $f_+(0)=1$. The deviation of $f_+(0)$
from unity represents the main source of theoretical uncertainty. This
deviation has been estimated many years ago by Leutwyler and Roos
(LR)\cite{LR}, who combined a leading order analysis in chiral perturbation
theory (ChPT) with a quark model calculation. They obtained $f_+^{K^0
\pi^-}(0) = 0.961 \pm 0.008$, and this value still represents the
referential estimate\cite{PDG}.

By averaging old experimental results for $K_{\ell 3}$ decays with the
recent measurement by E865 at BNL\cite{E865}, and using the LR 
determination of the vector form factor, the PDG quotes $|V_{us}| = 0.2200
\pm 0.0026$\cite{PDG}. This value, once combined with the determination of
$|V_{ud}|$ given in Eq.~(\ref{eq:vud}), implies about $2 \,\sigma$ deviation
from the CKM unitarity condition, i.e. $|V_{us}|^{\rm unit.}\simeq
\sqrt{1-|V_{ud}|^2} = 0.2265 \pm 0.0022$. 

\section{$K\ell 3$ decays: the new experimental results}
\label{sec:newexp}
\begin{figure}[t]
\begin{center}
\includegraphics[width=7.5cm]{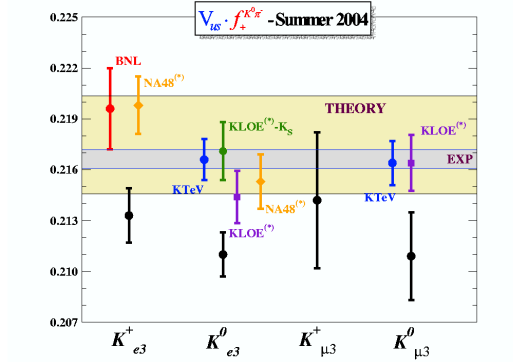}
\end{center}
\vspace{-0.5truecm}
\caption{\it Experimental results for $|V_{us}|\cdot f_+(0)$. The ``EXP"
and ``THEORY" bands indicate respectively the average of the new
experimental results and the unitarity prediction combined with the LR and
lattice (see Sect.\ref{sec:lattice}) determination of the vector form
factor.}
\label{fig:fvus}
\end{figure}

With respect to the PDG 2004 analysis, however, a significant novelty is
represented by several new experimental results, for both charged and
neutral $K_{\ell 3}$ decays, which have been recently presented by
KTeV\cite{KTEV}, NA48\cite{NA48} and KLOE\cite{KLOE}. Expressed in terms of
$|V_{us}|\cdot f_+(0)$, these determinations are shown in
Fig.~\ref{fig:fvus}, together with the BNL result and the averages of the
old $K_{\ell 3}$ results quoted by the PDG.
Remarkably, the average of the new results\cite{Mescia}, represented by
the darker band in the plot (``EXP"), is in very good agreement with the
unitarity prediction, once the LR determination of the vector form factor is
taken into account. The unitarity prediction is shown in Fig.~\ref{fig:fvus}
by the lighter band (``THEORY").

\section{$K\ell 3$ decays: theory status}
\label{sec:kl3}
An important theoretical progress in the determination of $|V_{us}|$ is
represented by the first (quenched) lattice determination, with significant
accuracy, of the vector form factor at zero momentum transfer
$f_+(0)$\cite{our}. The lattice result turns out to be in very good
agreement with the quark model estimate obtained by LR, thus putting the
evaluation of this form factor on a firmer theoretical basis. Before
outlining the strategy of the lattice calculation, I would like to summarize
the theoretical status of the $f_+(0)$ evaluations.

A good theoretical control on $K_{\ell 3}$ transitions is assured by the 
Ademollo-Gatto (AG) theorem\cite{ag}, which states that $f_+(0)$ is 
renormalized only by terms of at least second order in the breaking of 
SU(3)-flavor symmetry. Nevertheless, the error on the shift of $f_+(0)$
from unity represents not only the main source of theoretical uncertainty
but it also dominates the overall error in the determination of $|V_{us}|$.

The amount of SU(3) breaking due to light quark masses can be investigated 
within ChPT, by performing an expansion of the form $f_+(0) = 1 + f_2 + f_4
+ \ldots$, where $f_n = {\cal{O}}(p^n)={\cal{O}}[{M^n_{K,\pi}} /(4\pi
f_\pi)^n]$. Thanks to the AG theorem, the first non-trivial term in the
chiral expansion, $f_2$, does not receive contributions of local operators
appearing in the effective theory and can be computed unambiguously in terms
of $M_K$, $M_\pi$ and $f_\pi$ ($f_2 = -0.023$, in the $K^0 \to \pi^-$
case\cite{LR}). The higher-order terms of the chiral expansion involve
instead the coefficients of local chiral operators, that are difficult to
estimate. The quark model calculation by LR provides an estimate of the
next-to-leading correction $f_4$, and it is based on a general
parameterization of the SU(3) breaking structure of the pseudoscalar meson
wave functions. 

An important progress in this study is represented by the complete 
two-loop ChPT calculation of $f_4$, performed in Refs.\cite{post,BT}. In
Ref.\cite{BT}, the result has been written in the form
\be
f_4 = \Delta(\mu) - \frac{8}{F_\pi^4} \left[ C_{12}(\mu) + C_{34}(\mu) 
\right] \left(M_K^2-M_\pi^2\right)^2 ,
\label{eq:f4}
\ee
where $\Delta(\mu)$ is expressed in terms of chiral logs and the ${\cal
O}(p^4)$ low-energy constants, while the second term is the analytic one
coming from the ${\cal O}(p^6)$ chiral Lagrangian. As can be seen from
Eq.~(\ref{eq:f4}), this local contribution involves a single combination of
two (unknown) chiral coefficients entering the effective Lagrangian at
${\cal O}(p^6)$. In addition, the separation between non-local and local
contribution quantitatively depends on the choice of the renormalization
scale $\mu$, only the whole result for $f_4$ being scale independent. This
dependence is found to be large\cite{cnp}; for instance, at three typical
values of the scale one finds
\be
\label{eq:deltamu}
\Delta(\mu) = \left\{ 
\begin{array}{ccc}
0.031\,, & 0.015\,, & 0.004 \\
\mu = M_\eta & \mu = M_\rho & \mu = 1 \,\GeV
\end{array}
\right. \,.
\ee
An important observation by Bijnens and Talavera\cite{BT} is that the 
combination of low-energy constants entering $f_4$ could be in principle 
constrained by experimental data on the slope and curvature of the scalar
form factor. The required level of experimental precision, however, is far
from what is currently achieved. Thus, one is left with either the LR result
or other model dependent estimates of the local term in Eq.~(\ref{eq:f4}).
Recent attempts in this direction include the estimate by resonance
saturation obtained in Ref.\cite{cir2} and the dispersive analysis of
Ref.\cite{jamin}. The model results, however, are in disagreement within
each other. In addition, the large scale dependence of the ${\cal{O}}(p^6)$
loop calculation shown in Eq.~(\ref{eq:deltamu}) seems to indicate that the
error $\pm 0.010$ quoted in Refs.\cite{BT,cnp,jamin} might be
underestimated.\footnote{A different factorization between local and
non-local contributions has been considered in Ref.\cite{cir2}, which partly
reduces the dependence on the factorization scale.} For all these reasons, a
first principle lattice determination of the vector form factor is of great
phenomenological relevance.

\section{Strategy of the lattice calculation}
\label{sec:lattice}
The first lattice calculation of the vector form factor at zero momentum
transfer has been recently presented in Ref.\cite{our}. In order to reach
the challenging goal of about 1\% error on the lattice determination of
$f_+(0)$, a new strategy has been proposed and applied in the quenched
approximation. This strategy is based on three steps.

\subsubsection*{1) Precise evaluation of the scalar form factor $f_0(q^2)$
at $q^2=q_{\rm max}^2$}
\begin{figure}[t]
\begin{center}
\begin{tabular}{cc}
\includegraphics[width=5.0cm]{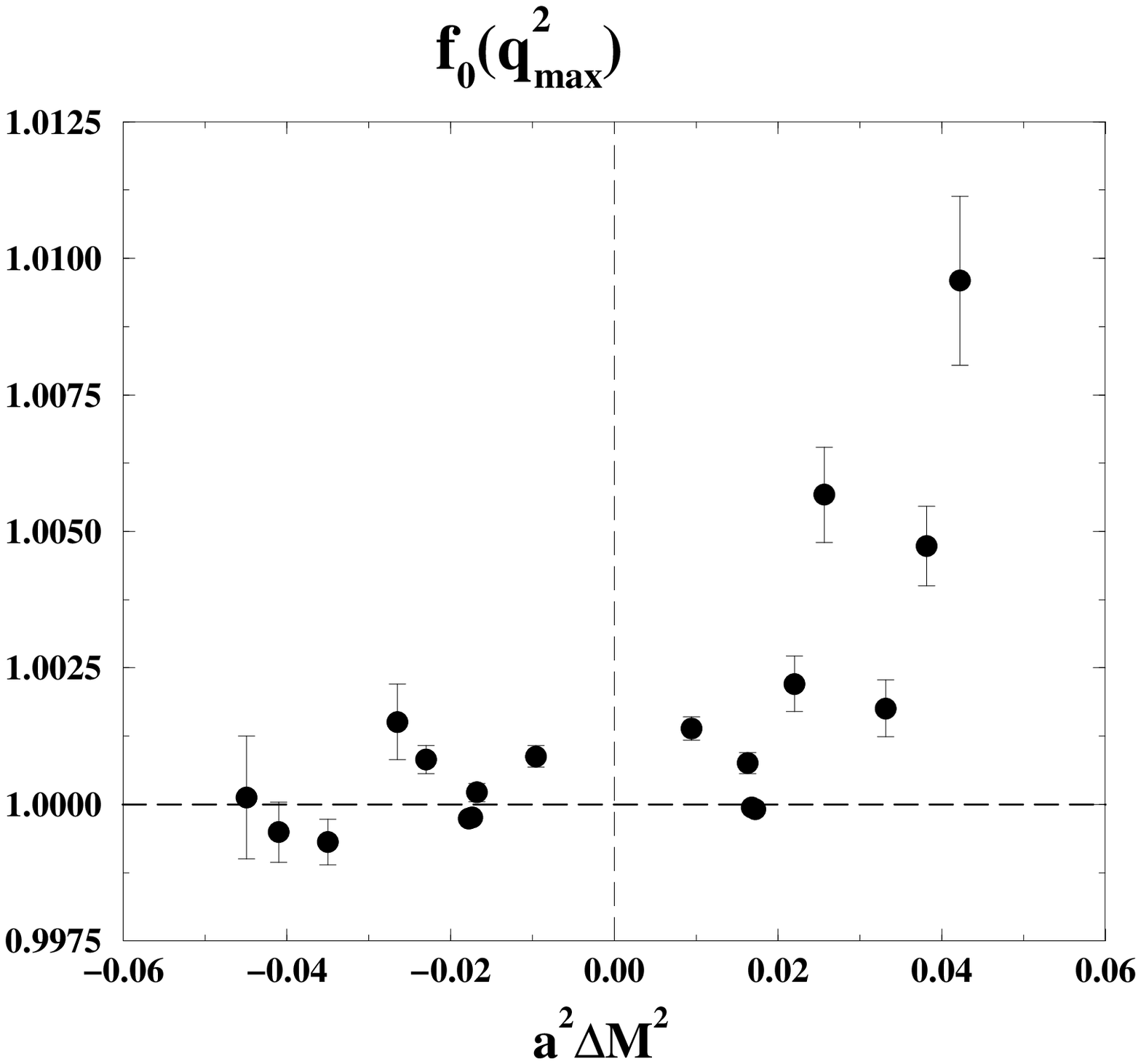} &
\includegraphics[width=5.3cm]{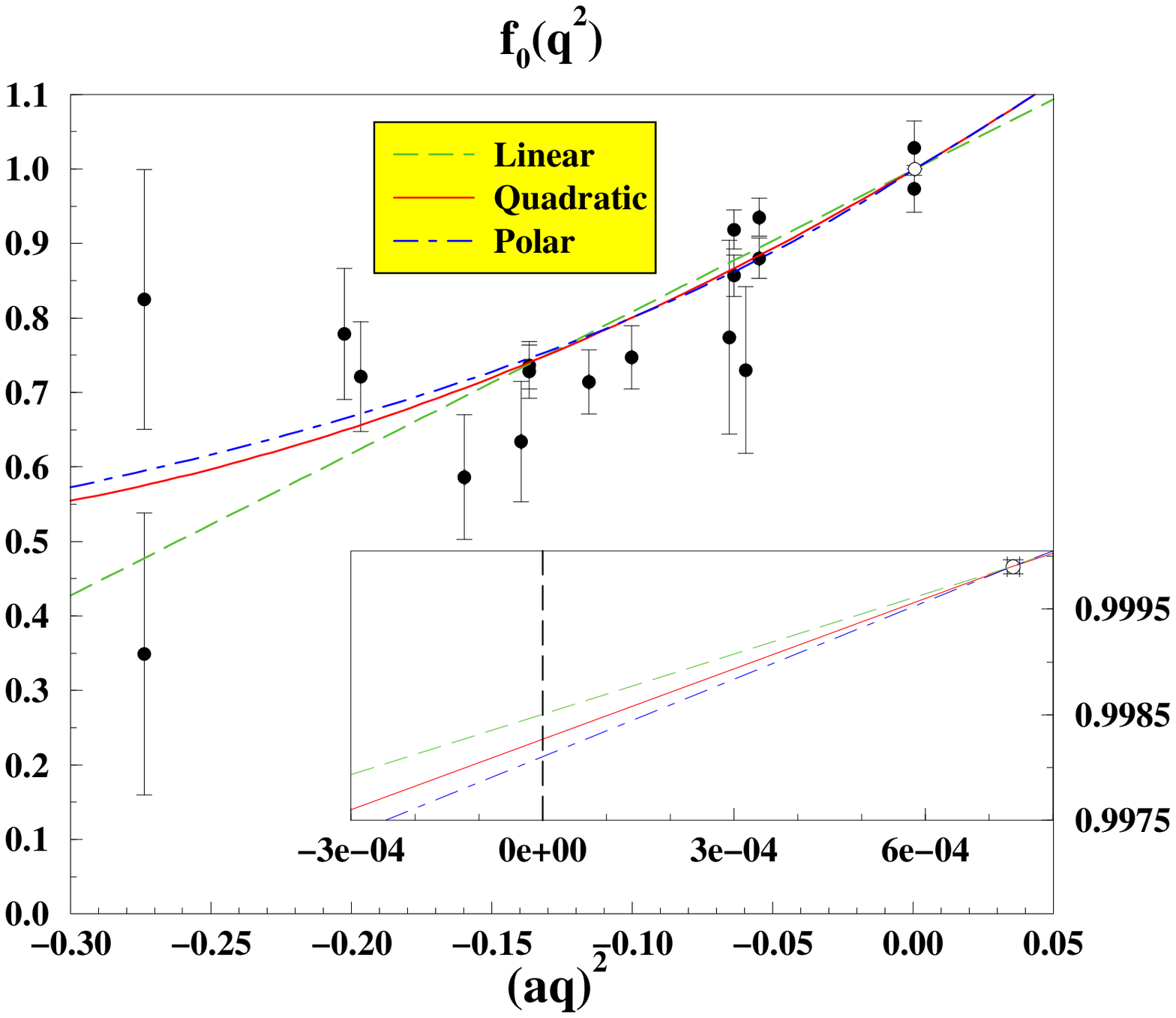}
\end{tabular}
\end{center}
\vspace{-0.5truecm}
\caption{\it {\rm Left:} Values of $f_0(q_{\protect\rm max}^2)$ versus the 
SU(3)-breaking parameter $a^2 \Delta M^2 \equiv a^2 (M_K^2 - M_{\pi}^2)$.
{\rm Right:} The form factor $f_0(q^2)$ as a function of $q^2$ for one of
the quark mass combinations. The inset is an enlargement of the region
around $q^2= 0$.}
\label{fig:f0}
\end{figure}

This evaluation follows a procedure originally proposed by the FNAL group
to study heavy-light form factors\cite{FNAL}. For $K_{\ell 3}$ decays, the
scalar form factor $f_0(q^2)$ can be calculated very efficiently at $q^2 =
q_{\rm max}^2 = (M_K - M_\pi)^2$ by studying the following double ratio of
matrix elements,
\be 
\frac{\langle \pi 
| \bar{s} \gamma_0 u | K \rangle \, \langle K | \bar{u} \gamma_0 s | 
\pi \rangle}{\langle K | \bar{s} \gamma_0 s | K \rangle \, \langle \pi | 
\bar{u} \gamma_0 u | \pi \rangle} = \frac{(M_K+M_\pi)^2}{4 M_K
M_\pi}\,[f_0(q_{\rm max}^2)]^2 \,,
\label{eq:fnal2}
\ee
where all the external particles are taken at rest. There are several
crucial advantages in using the double ratio (\ref{eq:fnal2}) which are
described in details in Ref.\cite{our}. The most important point is that
this ratio gives values of $f_0(q_{\rm max}^2)$ with a statistical
uncertainty smaller than 0.1\%, as it is illustrated in Fig.~\ref{fig:f0}
(left).

\subsubsection*{2) Extrapolation of $f_0(q_{\rm max}^2)$ to $f_0(0) =
f_+(0)$}
This extrapolation is performed by studying the $q^2$-dependence of
$f_0(q^2)$. New suitable double ratios are also introduced in this step,
that improve the statistical accuracy of $f_0(q^2)$. The quality of the
extrapolation is shown in Fig.~\ref{fig:f0} (right). Three different
functional forms in $q^2$ have been considered, namely a polar, a linear and
a quadratic one. The lattice result for the slope $\lambda_0$ of the scalar
form factor is in very good agreement with the recent accurate determination
from KTeV\cite{KTEV_slope}. In the case of the polar fit, for instance, the
lattice result is $\lambda_0 = 0.0122(22)$ (in units of $M_{\pi^+}^2$) to be
compared with the experimental determination $\lambda_0 = 0.0141(1)$.

\subsubsection*{3) Extrapolation to the physical masses}
\begin{figure}[t]
\begin{center}
\includegraphics[width=6.0cm]{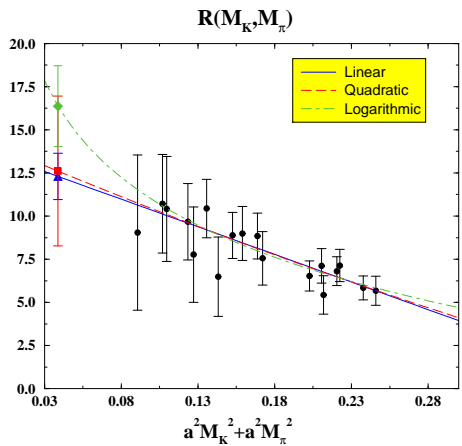}
\end{center}
\vspace{-0.5truecm}
\caption{\it Comparison among linear, quadratic and logarithmic fits
of the ratio $R(M_K, M_\pi)$ as a function of $[a^2 M_K^2 + a^2
M_{\pi}^2]$.}
\label{fig:fitcfr}
\end{figure}

The physical value of $f_+(0)$ is finally reached after extrapolating the 
lattice results to the physical kaon and pion masses. The problem of the
chiral extrapolation is substantially simplified if the AG theorem (which
holds also in the quenched approximation\cite{colangelo}) is taken into
account and if the leading (quenched) chiral logs are subtracted. This is
achieved by introducing the following ratio
\be
R = \frac{\Delta f}{(M_K^2-M_\pi^2)^2} = \frac{ 1 + f_2^q -f_+(0) }
{(M_K^2-M_\pi^2)^2} \,,
\label{eq:linfit}
\ee
where $f_2^q$ represents the leading chiral contribution calculated in
quenched ChPT\cite{our} and the quadratic dependence on $(M_K^2-M_\pi^2)$,
driven by the AG theorem, is factorized out. It should be emphasized that
the subtraction of $f_2^q$ in Eq.~(\ref{eq:linfit}) does not imply
necessarily a good convergence of (quenched) ChPT at ${\cal O}(p^4)$ for
the meson masses used in the lattice simulation. The aim of the subtraction
is to access directly on the lattice the quantity $\Delta f$, defined in
Eq.~(\ref{eq:linfit}) in such a way that its chiral expansion starts at
${\cal O}(p^6)$ independently of the values of the meson masses. After the
subtraction of $f_2^q$, the ratio $R$ of Eq.~(\ref{eq:linfit}) is smoothly
extrapolated in the meson masses as illustrated in Fig.~\ref{fig:fitcfr}. 
In order to check the stability of the results, linear, quadratic and
logarithmic fits have been considered. The chiral extrapolation leads to the
final result
\be
f_+^{K^0\pi^-}(0) = 0.960 \pm 0.005_{\rm stat} \pm 0.007_{\rm syst}\,,
\label{eq:f0final}
\ee
where the systematic error does not include quenching effects beyond ${\cal{
O}}(p^4)$. Removing this error represents one of the major goal of future 
lattice studies of $K_{\ell3}$ decays. Remarkably, two preliminary
unquenched calculations have been already presented. The results
read\cite{oka,shoij} 
\begin{eqnarray}
&& f_+^{K^0\pi^-}(0) = 0.962 \pm 0.006 \pm 0.007 \\
&& f_+^{K^0\pi^-}(0) = 0.954 \pm 0.009 \,,
\label{eq:f0unq}
\end{eqnarray}
in very good agreement with the quenched estimate of Eq.~(\ref{eq:f0final}).

The value (\ref{eq:f0final}) compares well with the LR result $f_+^{K^0
\pi^-}(0) = 0.961 \pm 0.008$ quoted by the PDG\cite{PDG} and, once combined
with the average of the more recent experimental results, implies 
\be
|V_{us}|=0.2256 \pm 0.0022 \,,
\ee
in good agreement with the unitarity prediction. 

A strategy similar to the one discussed above has been also applied to
study hyperon semileptonic decays on the lattice, and preliminary results
have been presented in\cite{iperoni}.

\section{Conclusions}
\label{sec:conclusions}
We have discussed the most recent experimental and theoretical progresses
achieved in the determination of $V_{us}$ from semileptonic kaon decays. On
the theoretical side, the main novelty is represented by the first lattice
QCD calculation of the $K_{\ell 3}$ vector form factor at zero-momentum
transfer, $f_+(0)$. This calculation is the first one obtained by using a
non-perturbative method based only on QCD, except for the quenched
approximation. Once combined with the new measurements of kaon semileptonic
decays, the lattice result leads to a determination of $V_{us}$ in very good
agreement with the expectation based on the determination of $V_{ud}$ and
the unitarity of the CKM matrix.

\section*{Acknowledgments}
It is a pleasure to thank the organizers for their success in creating the
stimulating and very much enjoyable atmosphere of the Rencontres.

\end{document}